\documentclass[letter]{aa}
\usepackage[colorlinks=true,citecolor=gray,linkcolor=blue]{hyperref}
\usepackage{siunitx}
\begin{document} 

\title{The shapes of column density PDFs}
\subtitle{The importance of the last closed contour}
\titlerunning{Completeness limit of PDFs} 
\author{Jo\~ao Alves\inst{1}, Marco Lombardi\inst{2}, Charles J.~Lada\inst{3}} \mail{joao.alves@univie.ac.at} 
\institute{University of Vienna, Department of Astrophysics, T\"urkenschanzstrasse 17, 1180 Vienna, Austria \and University of Milan, Department of Physics, via Celoria 16, I-20133 Milan, Italy \and Harvard-Smithsonian Center for Astrophysics, Mail Stop 72, 60 Garden Street, Cambridge, MA 02138} 
\date{Received ; Accepted  }

\abstract {The probability distribution function of column density (PDF) has become the tool of choice for cloud structure analysis and star formation studies. Its simplicity is attractive, and the PDF could offer access to cloud physical parameters otherwise difficult to measure, but there has been some confusion in the literature on the definition of its completeness limit and shape at the low column density end.   In this letter we use the natural definition of the completeness limit of a column density PDF, the last closed column density contour inside a surveyed region, and apply it to a set of large-scale maps of nearby molecular clouds. We conclude that there is no observational evidence for log-normal PDFs in these objects.  We find that all studied molecular clouds have PDFs well described by power laws, including the diffuse cloud Polaris.  Our results call for a new physical interpretation of the shape of the column density PDFs. We find that the slope of a cloud PDF is invariant to distance but not to the spatial arrangement of cloud material, and as such it is still a useful tool for investigating cloud structure. }


\maketitle
%

\section{Introduction}
\label{sec:introduction}

The probability distribution function (PDF) of column densities  in molecular clouds has become a popular tool for describing molecular cloud structure.  There is a broad consensus among observers and theorists that the PDF of molecular clouds are characterized by  a log-normal peak and by a ``power-law tail'' towards the high column-densities \cite[e.g.,][]{Lombardi2008,Kainulainen2009c,Goodman2009,Froebrich2010,Lombardi2010b,Schneider2011,Beaumont2012,Kainulainen2013,Alves2014,Schneider2015,Abreu-Vicente2015}. Log-normal PDFs for column density maps of turbulent clouds were first predicted from theoretical work \citep{Vazquez-Semadeni1994,Padoan1997,Scalo1998,Ostriker2001} and have been used as promising tools to compare numerical simulations with observations and theory \cite[e.g.,][]{Federrath2010,Renaud2013,Ward2014,Myers15,Donkov2017}. If observed, log-normal PDFs would offer tremendous insight into cloud physics, allowing  hard-to-measure parameters such as turbulent driving, Mach number, and magnetic pressure to be quantified \cite[e.g.,][]{Nordlund1999,Federrath2008,Federrath2010,Molina2012}. 

Recently, \cite{Lombardi15} investigated the effects of survey boundaries and cloud superposition on the column density PDF for eight molecular cloud complexes, and concluded that molecular clouds are, surprisingly, not well described by log-normal functions, but can  instead be described by power laws with exponents ranging from about $-4$ to $-2$. Also recently, \cite{Ossenkopf2016} investigated through simulations of synthetic clouds how the determination of PDFs is affected by noise, line-of-sight contamination, survey boundaries, and the incomplete sampling in interferometric measurements. \cite{Ossenkopf2016} conclude that inferences from the PDF parameters can be wrong by large factors, in particular at the low column density end of the PDF, but these authors fall short of ruling out the log-normal peak of the PDF as an artifact of the observations of molecular clouds. The conclusion in \cite{Ossenkopf2016} on the impact of survey boundaries was a call for surveys to cover at least 50\% of a cloud complex. While this is trivial to do in a controlled numerical experiment,  clouds have ill-defined boundaries and observationally it is  simply not possible to determine 50\% of an unknown cloud size. 

In this letter we use the natural definition of the completeness limit of a column density PDF, namely, the column density value at the last closed contour of column density inside a surveyed cloud area. To minimize
confusion, we apply this definition to a set of relatively high Galactic latitude molecular clouds using low-noise extinction-calibrated \textit{Herschel/Planck} emission data, and conclude that the well-known log-normal-looking peak of the PDF falls systematically below the PDF completeness limit. This indicates that the commonly observed log-normal-looking PDF peaks of molecular clouds do not represent the true PDF, but instead a convoluted consequence of data incompleteness and a variable background level.

\section{Completeness limit of a column density PDF}
\label{sec:data}

Measurements of column density PDFs of the ISM require the imposition of an observational boundary. This boundary can be dictated either by the spatial limits of an observational survey (the map borders) or by the need to separate different clouds inside the same survey. We show  that the imposition of a boundary significantly affects the completeness limit of the measured PDF at the low column density end. For example, consider a cloud represented by the black column density contours in the diagram in Figure 1. To make a measurement of the PDF of this cloud,  a boundary must be imposed within which the measurement is made (blue box). When the material extends beyond the boundary of the box, which is always the case observationally as there is no zero column density line of sight, the completeness is set by the value of the column density corresponding to the last closed contour of column density inside the box. Any value of column density below this last closed contour (i.e., any value on an open contour) will not be represented in a complete manner in the PDF of the cloud. 

Column densities represented by open contours are incomplete and cannot be corrected in a reliable way because an observer does not have access to information beyond the map boundaries. The last closed contour represents then the natural completeness limit of a cloud PDF, if above the nominal noise of the map and not affected by an unrelated cloud inside the same map. The concept of the last closed contour is not new in the literature;  in particular regarding column density PDFs, \cite{Kainulainen2011,Kainulainen2013} and \cite{Ossenkopf2016} point to it as the natural definition of completeness of a column density PDF. We note that for a real cloud, and depending on cloud structure, column density dynamic range, and map size, there may likely be more than one closed connected surface like the red connected surface in Figure 1. 

\begin{figure}
  \centering
  \includegraphics[width=\hsize]{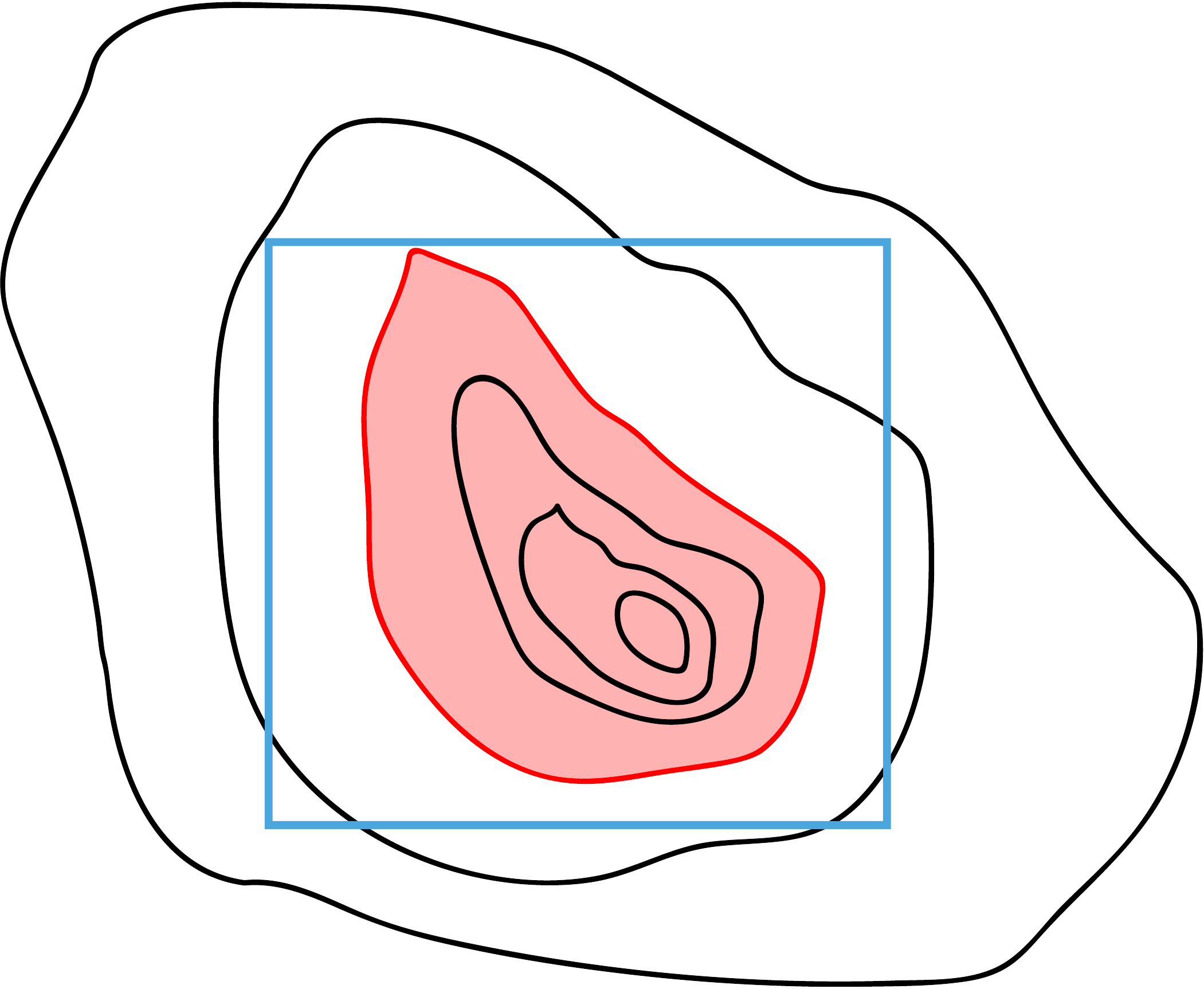}
  \caption{Survey (blue box) of a cloud (black column density contours). The PDF of this schematic survey is only complete to the last closed contour (red). The effects of noise, resolution, and superposition of clouds along the line of sight will only make the completeness more conservative than the red contour.}
  \label{fig:PDFcartoon}
\end{figure}

\begin{figure*}
  \centering
  \includegraphics[width=17cm]{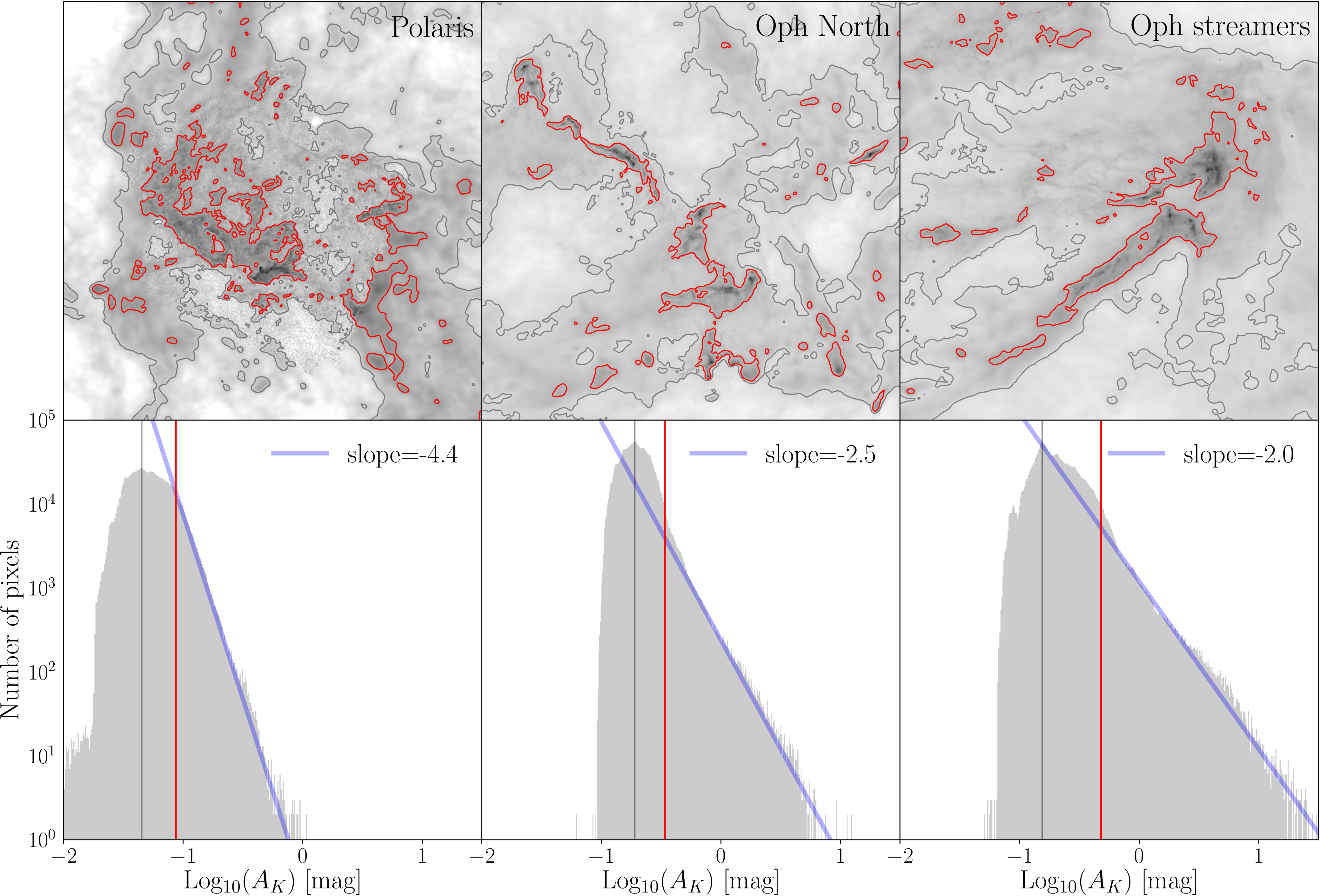}%
  \caption{Column density map and respective PDF for a diffuse cloud (Polaris), a star formation poor cloud (Oph North), and a star forming rich cloud (Oph streamers). The red line on the map and on the PDF corresponds to the last closed contour while the gray line corresponds to the peak of the PDF. The blue line represents the slope of the distribution above the completeness limit. All maps cover about the same physical area of about \SI{200}{pc^2}.}
  \label{mapsPDFs}
\end{figure*}
\section{Results}
\label{sec:results}

How does the boundary of a column density map affect the PDF? Guided by the simple idea behind the diagram in Figure~\ref{fig:PDFcartoon}, we investigate the implication of a last closed contour on real data and present three illustrative cases covering a wide range of possibilities from diffuse clouds to dense clouds that are star forming. In Figure~\ref{mapsPDFs} we present the map and the respective column density PDF of a) a diffuse and non-star forming cloud (Polaris), b) a cloud with low levels of star formation (Oph North), and c) an active, stellar cluster forming cloud (Ophiuchus Streamers). The column density maps, which are constructed from \textit{Herschel}, \textit{Planck}, and \textit{2MASS } data and are taken from \cite{Lombardi15}, cover about \SI{200}{pc^2} for each of these nearby clouds with a resolution of about \SI{0.02}{pc}. 

Figure~\ref{mapsPDFs} illustrates well the effect the map boundaries impose on the PDF. The maps in this figure have two contours, a gray contour representing  the column density level at the peak of the PDF and a red contour representing the  column density level at the last closed contour. These contours are represented by vertical lines on the respective PDF (the gray and red vertical lines). It is immediately clear from Figure~\ref{mapsPDFs} that the ``log-normal-looking'' peaks of the three PDFs falls below the respective completeness limits, and that the PDFs of the three sampled clouds are fairly well characterized by power laws. We tested that the shape of the complete PDF is not sensitive to small variations in the definition of the last closed contour for a map. The values of the power-laws fits above the completeness limit presented in this figure were estimated using  the Markov Chain Monte Carlo (MCMC) algorithm \texttt{emcee} \cite{Foreman-Mackey2013} on the logarithmically binned column densities. We find that different subregions of a particular map have different PDFs, in a manner similar to \cite{Stutz2015}, so the power-laws slopes presented here should be seen as representative of the entire cloud material inside the maps and above the respective column density completeness contour. Figure~\ref{mapsPDFs} gives another insight into what is happening below the last closed contour completeness limit. The peak of the distribution below the completeness limit is more pronounced for the Ophiuchus clouds than for Polaris, which reflects the different background level for these two different lines of sight. The Polaris cloud lies at $26^\circ$ above the Galactic plane against a much ``cleaner'' background, while the Ophiuchus clouds lie against a much more complex background. 

\section{Discussion}
\label{sec:discussion}

The results in this letter are in tension with the commonly accepted view that column density PDFs of molecular clouds are well described by a log-normal, or a log-normal with a power-law tail at the high column density end. Instead, our results clearly indicate that the PDFs of these clouds are simple power laws down to their completeness limit, regardless of the star formation activity of the cloud. The current interpretation for the shape of molecular cloud PDFs, where the log-normal peak is seen as a consequence of super-sonic turbulence (but see \citealt{Tassis2010}) and the power-law tail as a consequence of gravity dominated regions in a turbulent cloud (e.g., \citealt{2011MNRAS.416.1436B}) is severely challenged: not only do we  not find  log-normal peaks, we find that diffuse and star forming clouds are both well described by power-law PDFs. A new physical interpretation for the shapes of molecular clouds PDFs is needed.

One can safely predict that the power laws that characterize these PDFs are not pure power laws in the sense that they will not extend over all extinction ranges. Apart from the minor deviations seen in Figure~\ref{mapsPDFs}, we expect a departure at the high column densities where the cloud column density will reach a maximum, and a departure at the lowest column density (the cloud is finite). The latter departure is likely to take place below the lowest completeness limit we were able to reach, namely A$_K<0.09$ (or A$_V<0.8$) mag for the Polaris cloud. 

From dust emission and extinction data alone, even for  higher Galactic latitude clouds such as the ones presented here, we have not detected a hint of a break in the power-law PDF at the low column densities.  Larger surveys would be able in principle to establish more complete PDFs, but there is an unexpected observational limitation: because molecular clouds do not have well-defined boundaries, and are distributed around the Galactic plane, larger surveys will necessarily contain interloper clouds that would contaminate the PDF. An obvious strategy would be to study even higher Galactic latitude clouds, but a quick search for these does not offer many good candidates, if any.

Alternatively, observations of atomic hydrogen (HI) could be attempted to measure the PDF below $A_V \leq \SI{1}{mag}$, as was done recently by \cite{Burkhart2015,Imara2016} for nearby molecular clouds. They find narrow log-normal PDFs for $A_V< \SI{1}{mag}$, but this is a difficult measurement to make, as explained in these papers, and some uncertainty remains on 1) how much  this PDF shape, particularly at the higher column densities, is affected by HI depletion at the HI-to-H$_2$ interface, and 2) at the lower column densities the PDF is unconstrained, due to the finite boundary of the region studied, which is smaller than the last closed contour for the lowest column densities measured. Determining the exact shape of the PDF below A$_V \leq 1$ mag is now the new frontier and more studies are needed.

\begin{figure}
  \centering
  \includegraphics[width=\hsize]{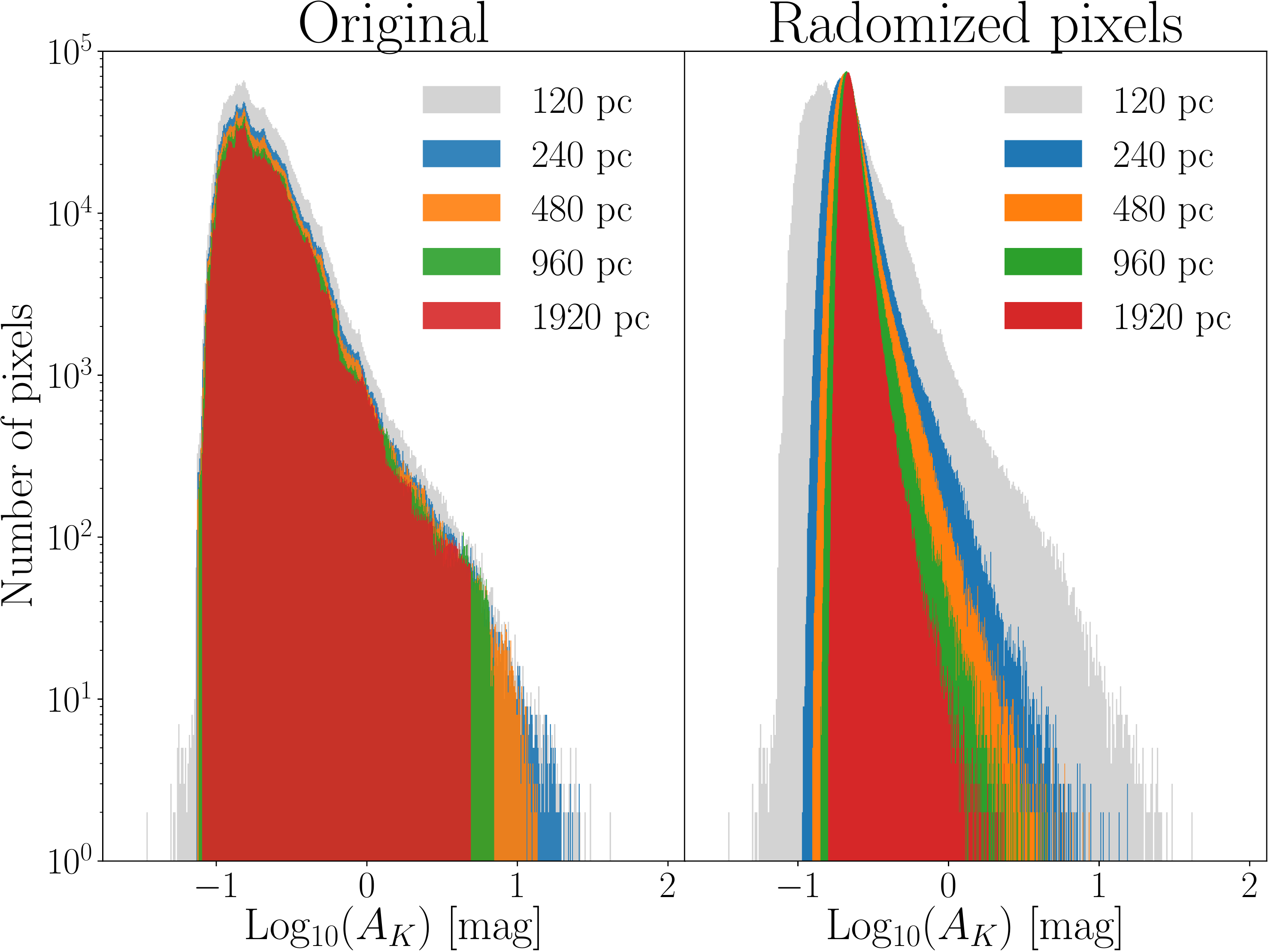}
  \caption{Left: Effect of distance on the Oph Streamers PDF. Right: Same, but with randomized pixel positions. The power-law PDF is invariant to distance (resolution), but not to the arrangement of cloud material.}
  \label{fig:PDFdistance}
\end{figure}

Column density PDFs are simplified 1-D representations of 2-D data sets. At first glance these PDFs have no information on the spatial distribution of cloud material\footnote{Chris Beaumont
  has perhaps the best visualization of this limitation, see \url{https://datarazzi.wordpress.com/2011/08/11/fractals-rho-ophiuchus-and-justin-bieber}}. A column density map, or a randomized distribution of pixels of the same map, will have the exact same PDF. While PDFs are fundamental image analysis tools, one may question their usefulness for cloud structure studies when they are apparently not sensitive to the spatial arrangement of cloud material. To test this contention we show in Figure~\ref{fig:PDFdistance} {(left)} the PDF of the Oph Streamers and {(right)} the PDF of a randomized distribution of pixels of the same Oph Streamers map. The PDFs are exactly the same for the distance of the cloud (120 pc, gray distribution), as expected. We then computed the PDF of the two maps, the original and the randomized one, for different distances (resolutions). Remarkably, as the distance (resolution) increases (decreases),  the power-law slope of the PDF of the original cloud remains essentially invariant to distance (resolution), while the PDF of the randomized cloud changes dramatically. This implies that the column density PDF is invariant to distance, but not to the spatial arrangement of cloud material, and as such it may ultimately prove to be a useful tool for investigating cloud structure.

Finally, the concept of last closed contour used here as the natural completeness limit for column density maps of molecular clouds should be applied, for the same reasons, to numerical work.

\section{Conclusions}
\label{sec:conclusions}

The main results of the paper can be summarized as follows:

\begin{enumerate}

\item The completeness limit for a column density PDF is defined by the last closed contour in a cloud's column density map. This is a best-case scenario that assumes no contamination from an unrelated cloud along the same line of sight, and a last closed contour above the nominal noise level of the map.

\item When this definition of completeness is applied to large-scale maps of molecular clouds, from diffuse clouds to star forming clouds, a consistent result appears: the well-known log-normal-looking peak falls systematically below the PDF completeness limit, indicating that it does not represent a feature of the true PDF of the cloud, but is a convoluted consequence of data incompleteness and a variable background level.

\item Our results call for a new physical interpretation of molecular clouds PDFs. We show that the log-normal-looking peak of the PDF is an artifact caused by data incompleteness, hence not a consequence of supersonic turbulence. Also, the interpretation of the power-law tail,  commonly taken as a consequence of gravity dominating over turbulence, is also in question as power laws describe both dense star forming clouds and diffuse clouds. 
  
\item The PDFs of these molecular clouds are well characterized by simple power laws, from maximum column density to the last closed contour. The power-law slope varies from cloud  to cloud, approximately from $-4$ (for diffuse clouds) to $-2$ (for star forming clouds), in agreement with \cite{Lombardi15}. Different subregions of a particular map have different PDFs, so these power laws should be seen as an average for an entire cloud. 

\item We find that the slope of a cloud PDF is invariant to distance but not to the spatial arrangement of cloud material, and as such it is a useful tool for investigating cloud structure.

\end{enumerate}

\begin{acknowledgements} 
  Based on observations obtained with \textit{Planck} and  \textit{Herschel}, ESA science missions with instruments and contributions directly funded by ESA Member States, NASA, and Canada. This research has made use of NASA's Astrophysics Data System. This research has made use of the SIMBAD database, operated at CDS, Strasbourg, France. This research has made use of the VizieR catalogue access tool, CDS, Strasbourg, France. This research has made use of Astropy, a community-developed core Python package for Astronomy \citep{2013A&A...558A..33A}. This research has made use of TOPCAT, an interactive graphical viewer and editor for tabular data \citep{2005ASPC..347...29T}.  
\end{acknowledgements}

\bibliographystyle{aa} 
\bibliography{zmy}

\begin{thebibliography}{34}
\expandafter\ifx\csname natexlab\endcsname\relax\def\natexlab#1{#1}\fi

\bibitem[{{Abreu-Vicente} {et~al.}(2015){Abreu-Vicente}, {Kainulainen},
  {Stutz}, {Henning}, \& {Beuther}}]{Abreu-Vicente2015}
{Abreu-Vicente}, J., {Kainulainen}, J., {Stutz}, A., {Henning}, T., \&
  {Beuther}, H. 2015, \aap, 581, A74

\bibitem[{{Alves} {et~al.}(2014){Alves}, {Lombardi}, \& {Lada}}]{Alves2014}
{Alves}, J., {Lombardi}, M., \& {Lada}, C.~J. 2014, \aap, 565, A18

\bibitem[{{Astropy Collaboration} {et~al.}(2013){Astropy Collaboration},
  {Robitaille}, {Tollerud}, {Greenfield}, {Droettboom}, {Bray}, {Aldcroft},
  {Davis}, {Ginsburg}, {Price-Whelan}, {Kerzendorf}, {Conley}, {Crighton},
  {Barbary}, {Muna}, {Ferguson}, {Grollier}, {Parikh}, {Nair}, {Unther},
  {Deil}, {Woillez}, {Conseil}, {Kramer}, {Turner}, {Singer}, {Fox}, {Weaver},
  {Zabalza}, {Edwards}, {Azalee Bostroem}, {Burke}, {Casey}, {Crawford},
  {Dencheva}, {Ely}, {Jenness}, {Labrie}, {Lim}, {Pierfederici}, {Pontzen},
  {Ptak}, {Refsdal}, {Servillat}, \& {Streicher}}]{2013A&A...558A..33A}
{Astropy Collaboration}, {Robitaille}, T.~P., {Tollerud}, E.~J., {et~al.} 2013,
  \aap, 558, A33

\bibitem[{{Ballesteros-Paredes} {et~al.}(2011){Ballesteros-Paredes},
  {V{\'a}zquez-Semadeni}, {Gazol}, {Hartmann}, {Heitsch}, \&
  {Col{\'{\i}}n}}]{2011MNRAS.416.1436B}
{Ballesteros-Paredes}, J., {V{\'a}zquez-Semadeni}, E., {Gazol}, A., {et~al.}
  2011, \mnras, 416, 1436

\bibitem[{{Beaumont} {et~al.}(2012){Beaumont}, {Goodman}, {Alves}, {Lombardi},
  {Rom{\'a}n-Z{\'u}{\~n}iga}, {Kauffmann}, \& {Lada}}]{Beaumont2012}
{Beaumont}, C.~N., {Goodman}, A.~A., {Alves}, J.~F., {et~al.} 2012, \mnras,
  423, 2579

\bibitem[{{Burkhart} {et~al.}(2015){Burkhart}, {Lee}, {Murray}, \&
  {Stanimirovi{\'c}}}]{Burkhart2015}
{Burkhart}, B., {Lee}, M.-Y., {Murray}, C.~E., \& {Stanimirovi{\'c}}, S. 2015,
  \apjl, 811, L28

\bibitem[{{Donkov} {et~al.}(2017){Donkov}, {Veltchev}, \&
  {Klessen}}]{Donkov2017}
{Donkov}, S., {Veltchev}, T.~V., \& {Klessen}, R.~S. 2017, \mnras, 466, 914

\bibitem[{{Federrath} {et~al.}(2008){Federrath}, {Klessen}, \&
  {Schmidt}}]{Federrath2008}
{Federrath}, C., {Klessen}, R.~S., \& {Schmidt}, W. 2008, \apjl, 688, L79

\bibitem[{{Federrath} {et~al.}(2010){Federrath}, {Roman-Duval}, {Klessen},
  {Schmidt}, \& {Mac Low}}]{Federrath2010}
{Federrath}, C., {Roman-Duval}, J., {Klessen}, R.~S., {Schmidt}, W., \& {Mac
  Low}, M.-M. 2010, \aap, 512, A81

\bibitem[{{Foreman-Mackey} {et~al.}(2013){Foreman-Mackey}, {Hogg}, {Lang}, \&
  {Goodman}}]{Foreman-Mackey2013}
{Foreman-Mackey}, D., {Hogg}, D.~W., {Lang}, D., \& {Goodman}, J. 2013, \pasp,
  125, 306

\bibitem[{{Froebrich} \& {Rowles}(2010)}]{Froebrich2010}
{Froebrich}, D. \& {Rowles}, J. 2010, \mnras, 406, 1350

\bibitem[{{Goodman} {et~al.}(2009){Goodman}, {Pineda}, \&
  {Schnee}}]{Goodman2009}
{Goodman}, A.~A., {Pineda}, J.~E., \& {Schnee}, S.~L. 2009, \apj, 692, 91

\bibitem[{{Imara} \& {Burkhart}(2016)}]{Imara2016}
{Imara}, N. \& {Burkhart}, B. 2016, \apj, 829, 102

\bibitem[{{Kainulainen} {et~al.}(2011){Kainulainen}, {Beuther}, {Banerjee},
  {Federrath}, \& {Henning}}]{Kainulainen2011}
{Kainulainen}, J., {Beuther}, H., {Banerjee}, R., {Federrath}, C., \&
  {Henning}, T. 2011, \aap, 530, A64

\bibitem[{{Kainulainen} {et~al.}(2009){Kainulainen}, {Beuther}, {Henning}, \&
  {Plume}}]{Kainulainen2009c}
{Kainulainen}, J., {Beuther}, H., {Henning}, T., \& {Plume}, R. 2009, \aap,
  508, L35

\bibitem[{{Kainulainen} {et~al.}(2013){Kainulainen}, {Federrath}, \&
  {Henning}}]{Kainulainen2013}
{Kainulainen}, J., {Federrath}, C., \& {Henning}, T. 2013, \aap, 553, L8

\bibitem[{{Lombardi} {et~al.}(2010){Lombardi}, {Alves}, \&
  {Lada}}]{Lombardi2010b}
{Lombardi}, M., {Alves}, J., \& {Lada}, C.~J. 2010, \aap, 519, L7

\bibitem[{{Lombardi} {et~al.}(2015){Lombardi}, {Alves}, \& {Lada}}]{Lombardi15}
{Lombardi}, M., {Alves}, J., \& {Lada}, C.~J. 2015, \aap, 576, L1

\bibitem[{{Lombardi} {et~al.}(2008){Lombardi}, {Lada}, \&
  {Alves}}]{Lombardi2008}
{Lombardi}, M., {Lada}, C.~J., \& {Alves}, J. 2008, \aap, 489, 143

\bibitem[{{Molina} {et~al.}(2012){Molina}, {Glover}, {Federrath}, \&
  {Klessen}}]{Molina2012}
{Molina}, F.~Z., {Glover}, S.~C.~O., {Federrath}, C., \& {Klessen}, R.~S. 2012,
  \mnras, 423, 2680

\bibitem[{{Myers}(2015)}]{Myers15}
{Myers}, P.~C. 2015, ArXiv e-prints

\bibitem[{{Nordlund} \& {Padoan}(1999)}]{Nordlund1999}
{Nordlund}, {\AA}.~K. \& {Padoan}, P. 1999, in Interstellar Turbulence, ed.
  J.~{Franco} \& A.~{Carraminana}, 218

\bibitem[{{Ossenkopf} {et~al.}(2016){Ossenkopf}, {Csengeri}, {Schneider},
  {Federrath}, \& {Klessen}}]{Ossenkopf2016}
{Ossenkopf}, V., {Csengeri}, T., {Schneider}, N., {Federrath}, C., \&
  {Klessen}, R.~S. 2016, ArXiv e-prints

\bibitem[{{Ostriker} {et~al.}(2001){Ostriker}, {Stone}, \&
  {Gammie}}]{Ostriker2001}
{Ostriker}, E.~C., {Stone}, J.~M., \& {Gammie}, C.~F. 2001, \apj, 546, 980

\bibitem[{{Padoan} {et~al.}(1997){Padoan}, {Nordlund}, \& {Jones}}]{Padoan1997}
{Padoan}, P., {Nordlund}, A., \& {Jones}, B.~J.~T. 1997, \mnras, 288, 145

\bibitem[{{Renaud} {et~al.}(2013){Renaud}, {Bournaud}, {Emsellem}, {Elmegreen},
  {Teyssier}, {Alves}, {Chapon}, {Combes}, {Dekel}, {Gabor}, {Hennebelle}, \&
  {Kraljic}}]{Renaud2013}
{Renaud}, F., {Bournaud}, F., {Emsellem}, E., {et~al.} 2013, \mnras, 436, 1836

\bibitem[{{Scalo} {et~al.}(1998){Scalo}, {Vazquez-Semadeni}, {Chappell}, \&
  {Passot}}]{Scalo1998}
{Scalo}, J., {Vazquez-Semadeni}, E., {Chappell}, D., \& {Passot}, T. 1998,
  \apj, 504, 835

\bibitem[{{Schneider} {et~al.}(2011){Schneider}, {Bontemps}, {Simon},
  {Ossenkopf}, {Federrath}, {Klessen}, {Motte}, {Andr{\'e}}, {Stutzki}, \&
  {Brunt}}]{Schneider2011}
{Schneider}, N., {Bontemps}, S., {Simon}, R., {et~al.} 2011, \aap, 529, A1+

\bibitem[{{Schneider} {et~al.}(2015){Schneider}, {Ossenkopf}, {Csengeri},
  {Klessen}, {Federrath}, {Tremblin}, {Girichidis}, {Bontemps}, \&
  {Andr{\'e}}}]{Schneider2015}
{Schneider}, N., {Ossenkopf}, V., {Csengeri}, T., {et~al.} 2015, \aap, 575, A79

\bibitem[{{Stutz} \& {Kainulainen}(2015)}]{Stutz2015}
{Stutz}, A.~M. \& {Kainulainen}, J. 2015, \aap, 577, L6

\bibitem[{{Tassis} {et~al.}(2010){Tassis}, {Christie}, {Urban}, {Pineda},
  {Mouschovias}, {Yorke}, \& {Martel}}]{Tassis2010}
{Tassis}, K., {Christie}, D.~A., {Urban}, A., {et~al.} 2010, \mnras, 408, 1089

\bibitem[{{Taylor}(2005)}]{2005ASPC..347...29T}
{Taylor}, M.~B. 2005, in Astronomical Society of the Pacific Conference Series,
  Vol. 347, Astronomical Data Analysis Software and Systems XIV, ed.
  P.~{Shopbell}, M.~{Britton}, \& R.~{Ebert}, 29

\bibitem[{{Vazquez-Semadeni}(1994)}]{Vazquez-Semadeni1994}
{Vazquez-Semadeni}, E. 1994, \apj, 423, 681

\bibitem[{{Ward} {et~al.}(2014){Ward}, {Wadsley}, \& {Sills}}]{Ward2014}
{Ward}, R.~L., {Wadsley}, J., \& {Sills}, A. 2014, \mnras, 445, 1575

\end{thebibliography}

\end{document}